\documentclass[journal]{IEEEtran}
\usepackage{cite}     
\usepackage{graphicx}
\usepackage{amsmath}

\begin{document}
\pagestyle{empty}
\title{Operational Experience and First Results with a Highly Granular Tungsten Analog Hadron Calorimeter}
\author{Frank Simon, on behalf of the CALICE Collaboration 
\thanks{F.~Simon is with the Max-Planck-Institut f\"ur Physik, Munich, Germany and with the Excellence Cluster Universe, Technical University Munich, Germany. ({\it email: fsimon@mpp.mpg.de}).} }

\maketitle

\begin{abstract}
Precision physics at future multi-TeV lepton colliders such as CLIC requires excellent jet energy resolution. The detectors need deep calorimeter systems to limit the energy leakage also for very highly energetic particles and jets. At the same time, compact physical dimensions are mandatory to permit the installation of the complete calorimeter system inside high-field solenoidal magnets. This requires very dense absorbers, making tungsten a natural choice for hadron calorimeters at such a future collider. To study the performance of such a calorimeter, a physics prototype with tungsten absorbers and scintillator tiles with SiPM readout as active elements has been constructed and has been tested in particle beams at CERN over a wide energy range from 1 GeV to 300 GeV. We report on the construction and on the operational experience obtained with muon, electron and hadron beams.
\end{abstract}

\thispagestyle{empty}

\section{Introduction}

The physics at future high-energy lepton colliders requires jet energy reconstruction with unprecedented precision. Detector concepts for the International Linear Collider (ILC) \cite{Brau:2007zza} and the Compact Linear Collider (CLIC) \cite{Assmann:2000hg} rely on Particle Flow Algorithms \cite{Brient:2002gh, pfaMorgunov, Thomson:2009rp} to achieve the necessary precision. This event reconstruction technique requires highly granular calorimeters to deliver optimal performance. Such calorimeters are developed and studied by the CALICE collaboration. 

At CLIC, with a center of mass energy of up to 3 TeV, the depth of the hadronic calorimeter system is of key importance to limit the impact of energy leakage also for highly energetic particles. Simulation studies have shown that a depth of at least 7.5 $\lambda_I$, in addition to a 1 $\lambda_I$ deep electromagnetic calorimeter, is necessary to achieve the required jet energy resolution. Since the complete calorimeter system has to be located inside the coil of the main solenoid of the experiments to avoid un-instrumented regions which deteriorate the energy resolution, compact detector designs are mandatory, in particular in the barrel region. To satisfy the space constraints while providing the necessary depth for adequate shower containment, tungsten is being considered as the absorber material for the barrel hadron calorimeter for detector systems at CLIC.

\section{The CALICE Tungsten Analog Hadron Calorimeter}

\begin{figure}
\centering
\includegraphics[width=0.45\textwidth]{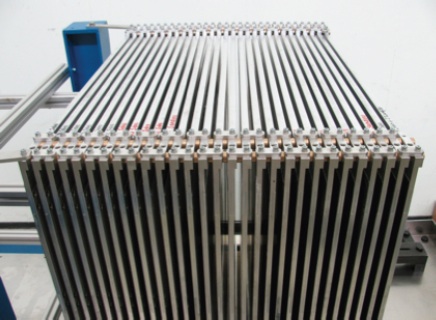}
\caption{The tungsten absorber structure, in the 30 layers configuration used during 2010 test beams.}
\label{fig:AbsorberStructure}
\end{figure}

To investigate the performance of hadron calorimeters with tungsten absorbers, and to gain first experience in the construction of such a detector, a physics prototype, the Tungsten Analog Hadron Calorimeter WHCAL, has been constructed. The absorber structure consists of a total of 39 tungsten layers, of which 30 were available in a first test beam period at the CERN PS in fall 2010.  The full 39 layer absorber structure with space for 38 active layers was used in beam tests with energies up to 300 GeV at the CERN SPS in 2011. The detector has a lateral size of approximately $81\,\times\,81\,\mathrm{cm}^2$. Each absorber plate consists of 10 mm tungsten alloy (93\% W, 1.8\% Cu, 5.2\% Ni, density 17.6 g/cm$^3$) and a 0.5 mm steel sheet which provides additional mechanical stability for the tungsten which is assembled from nine smaller tiles with an edge length of 27 cm. Figure \ref{fig:AbsorberStructure} shows the absorber structure prior to the installation of the active detectors. 

The active elements of the detector are taken from the CALICE analog hadron calorimeter with steel absorber \cite{Adloff:2010hb}, which has been extensively tested in particle beams at DESY, CERN and Fermilab. These active layers consist of small plastic scintillator tiles with sizes ranging from $3\,\times\,3\,\mathrm{cm}^2$ to $12\,\times\,12\,\mathrm{cm}^2$, each read out by a  Silicon Photomultipiers (SiPMs)\cite{Bondarenko:2000in} coupled to a wavelength-shifting fiber embedded in the scintillator. The scintillator cells are housed in cassettes with 2 mm thick steel front and back covers, resulting in a total of 4.5 mm steel and 10 mm tungsten alloy per layer, giving approximately 3.8 $\lambda_I$ of depth for the 30 layer configuration of the detector and a depth of 4.8 $\lambda_I$ in the 38 layer configuration used for energies above 10 GeV at the SPS. 

To control leakage at the highest energies, a tail catcher and muon tracker using 16 layers of scintillator strips with SiPM readout with an orientation alternating from layer to layer. The active elements are taken from the CALICE TCMT \cite{Dyshkant:2006et} used in test beam campaigns at CERN and at Fermilab from 2006 to 2011. The absorber structure has been newly constructed since the original absorber structure was used for tests of a digital hadron calorimeter \cite{Repond:2011jc} at Fermilab.

\section{Data Set and Detector Performance}

The 30 layer WHCAL has been exposed to mixed particle beams at the CERN PS, containing muons, electrons, pions and protons at momenta ranging from 1 GeV/c to 10 GeV/c. In total, 28 million events have been recorded. 

The full 38 layer detector has been exposed to particle beams at the CERN SPS in summer 2011, with a total of 33 million events recorded, collecting electrons in the range from 10 GeV to 40 GeV, charged pions and mixed hadron beams at energies from 10 GeV to 300 GeV, 180 GeV muons for calibrations and a special sample of tagged kaons. 
 
The  lower beam energies are particularly important for the testing of various Geant4 physics lists \cite{Geant4Physics}, which often have transition regions between different models at energies of a few GeV. The highest energies provide crucial information on the performance of a tungsten HCAL at single particle energies relevant for jets at CLIC. 

\begin{figure}
\centering
\includegraphics[width=0.45\textwidth]{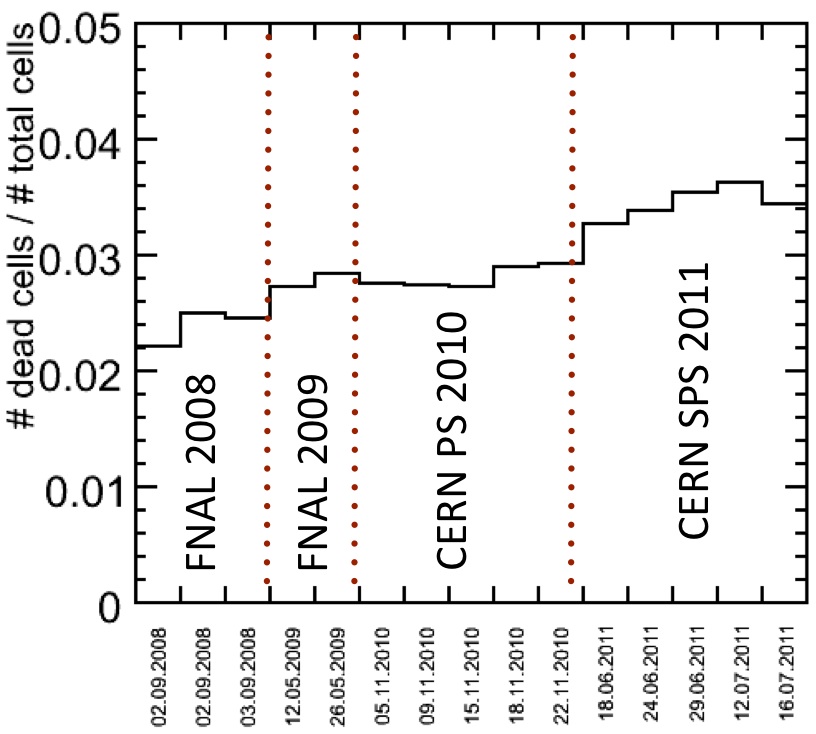}
\caption{Evolution of non-working channels of the AHCAL modules over several test beam periods.}
\label{fig:DeadChannels}
\end{figure}
 
Approximately two times per day during data taking the gain of the photon sensors is determined using a built-in LED calibration system \cite{Adloff:2010hb}, which also allows to monitor the number of defective electronics channels. Figure \ref{fig:DeadChannels} shows the evolution of the fraction of non-working channels of the AHCAL modules over several test beam periods, starting with data-taking at Fermilab with steel absorbers in fall 2008 and extending to the second-to-last beam period at the CERN SPS with tungsten absorbers.  Between the FNAL and the CERN periods, the calorimeter was completely disassembled and transported by truck and ship from the US to Europe, first to Hamburg for testing, and then to CERN for the test beam campaigns. Between the CERN PS and SPS periods, the calorimeter was again completely disassembled and moved from the PS East Area to the SPS North Area. The defective channels are mostly due to broken solder connections of the photon sensors, and not due to failures of the photon sensors themselves. Overall, the fraction of non-working channels is very low, considering the six years of test beam operation at DESY, CERN and Fermilab and the numerous assembly and disassembly procedures and the transportation of the modules, demonstrating the robustness of the detector technology and of the mechanical design. 

\begin{figure}
\centering
\includegraphics[width=0.45\textwidth]{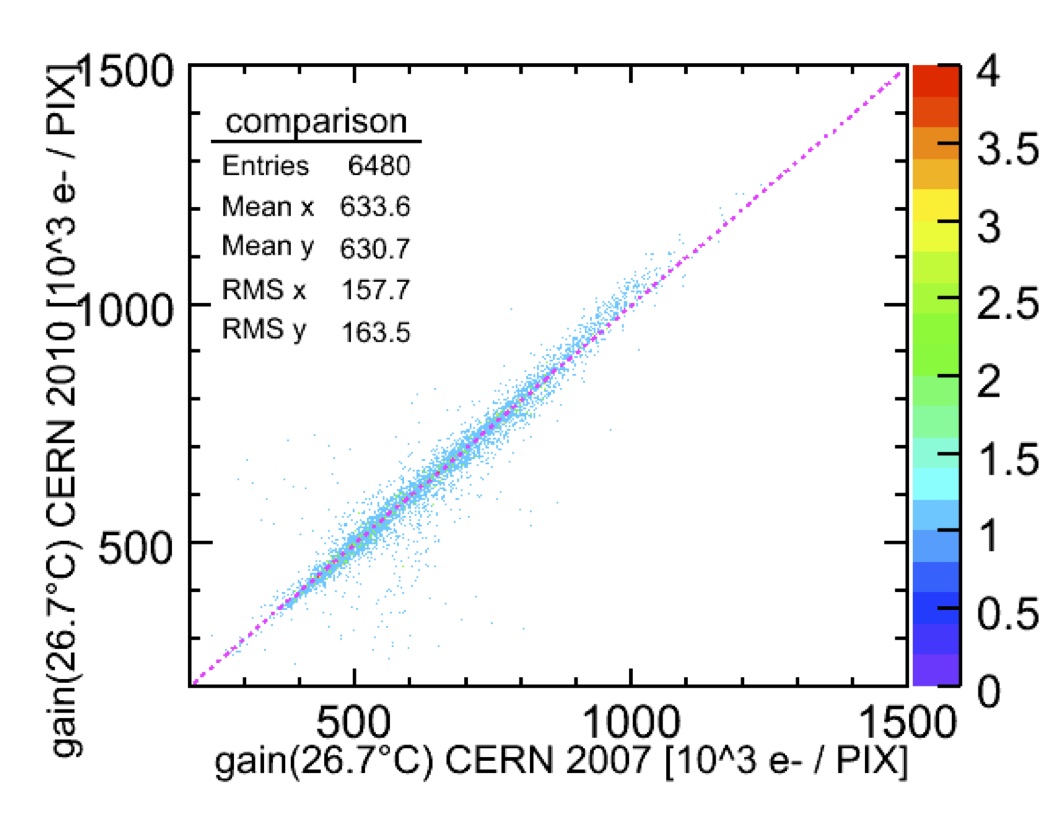}
\caption{Correlation of gain values determined in 2007 and 2010, corrected for temperature effects.}
\label{fig:CalibrationCorrelation}
\end{figure}

The reproducibility and stability of the gain calibration values of the photon sensors is studied by comparing  the values determined at CERN in 2007 with the ones obtained in 2010. The correlation of the two measurements is shown in Figure \ref{fig:CalibrationCorrelation}. Since the gain of the photon sensors is temperature dependent, both measurements are corrected to one common temperature using temperature coefficients determined over a wide temperature range during the test beam campaigns. The high correlation observed between these two measurements taken three years apart shows the stability of the photon sensors, and the overall quality of the gain calibration and its temperature correction.

\section{First Results} 

\begin{figure}
\centering
\includegraphics[width=0.45\textwidth]{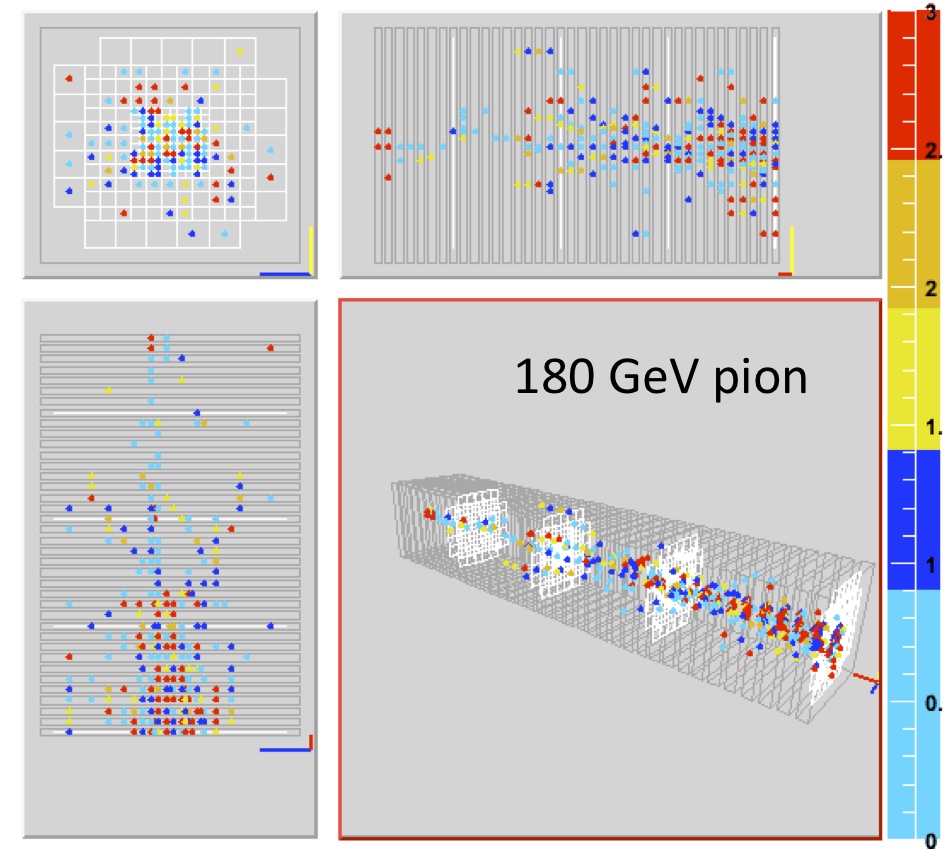}
\caption{Event display of a 180 GeV negative pion. The color scale shows the energy deposits per cell in units of minimum-ionizing particles.}
\label{fig:EventDisplay}
\end{figure}

Figure \ref{fig:EventDisplay} shows an event display of the shower of a \mbox{180 GeV} negative pion in the WHCAL. The display illustrates the imaging capabilities of the calorimeter, and shows compact electromagnetic subshowers with high energy density as well as extended, sparser hadronic components. 
 
The energy calibration of the response of the individual calorimeter cells is based on muons, using the energy deposit of a minimum-ionizing particle as reference scale. For a first analysis of the data, calibration constants for the scintillator cells from the 2007 test beam campaign with the CALICE steel AHCAL at the CERN SPS and preliminary values obtained with low-energy muons from the CERN PS East Hall are used. Both of these techniques lead to increased uncertainties and reduced energy resolution. At the PS, neither the angle nor the momentum of the muons was well defined, with a spread of approximately $\pm$ 20$^\circ$ in angle, and muon energies ranging from 1 GeV to 10 GeV. This results in an increased spread of the energy loss of the particles in the scintillator layers, and a correspondingly increased calibration uncertainty. A full calibration data set has been recorded during the test beam campaign at the CERN SPS in 2011 and is at present being analyzed, to be used in the future analysis of the electron and hadron data. 

\begin{figure}
\centering
\includegraphics[width=0.45\textwidth]{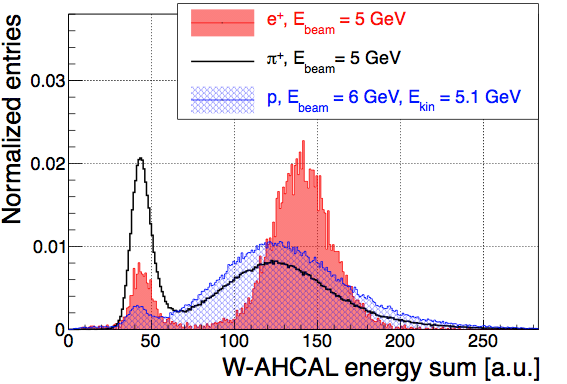}
\caption{Reconstructed energies for muons, positrons, $\pi^+$ and protons with kinetic energies around 5 GeV.}
\label{fig:5GeVReco}
\end{figure}

Figure \ref{fig:5GeVReco} shows the distribution of reconstructed energies for positrons, muons and pions as well as protons with kinetic energies around 5 GeV. Proton - pion separation is provided by threshold Cherenkov counters in the beam line, while muons and pions can only be partially separated with the Cherenkov counters due to in-flight decays and cross efficiencies. Above \mbox{3 GeV}, muons are well separated from the other particles due to their reduced energy deposition in the calorimeter. At PS energies, below 10 GeV, the response to electrons is higher than that to hadrons, indicating an $e/\pi$ ratio greater than 1 in this energy range.

\begin{figure}
\centering
\includegraphics[width=0.45\textwidth]{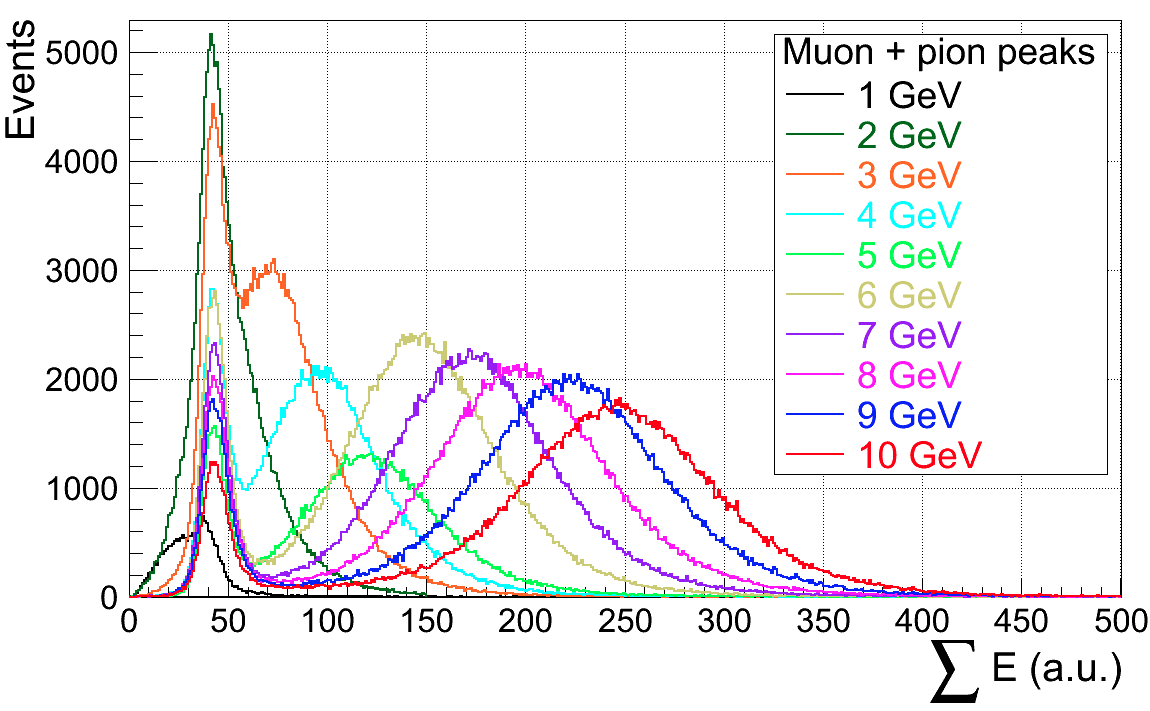}
\caption{Reconstructed energy for mixed muon and pion samples at momenta from 1 GeV/c to 10 GeV/c recorded at the CERN PS.}
\label{fig:Energies}
\end{figure}

Figure \ref{fig:Energies} shows the distribution of reconstructed energies for mixed muon and pion samples at momenta from 1 GeV/c to \mbox{10 GeV/c}. Below 3 GeV, the separation of muon and pion signals is possible by means of combined fits which provide mean and width for the individual components. At higher energies, the values can be extracted with simple fits to the respective peaks. 

\begin{figure}
\centering
\includegraphics[width=0.45\textwidth]{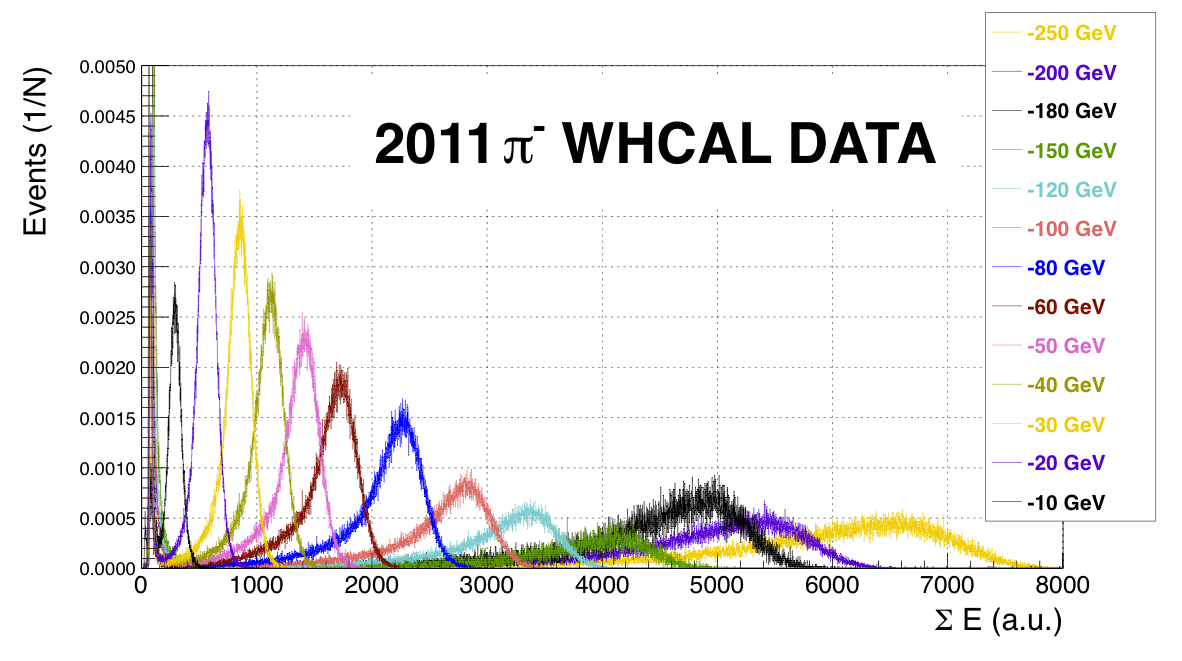}
\caption{Reconstructed energy for negative pion samples at momenta from 10 GeV/c to 250 GeV/c recorded at the CERN SPS.}
\label{fig:HighEnergies}
\end{figure}

Figure \ref{fig:HighEnergies} shows the distribution of reconstructed energies in the tungsten HCAL for the higher energy sample, ranging from 10 GeV/c to 250 GeV/c. At energies above 30 GeV, leakage starts to play an increasingly important role due to the increasing depth of the showers, shown by the tail to lower reconstructed energies present in the distributions. In the further analysis of the data sample, this will be controlled by the information provided by the tail catcher, which extends the depth of the calorimeter system by an additional 5.2 $\lambda_I$. The effect of leakage can also be largely eliminated by requiring a shower start in the first two layers of the calorimeter, which can be identified using the high granularity of the detector.

\section{Outlook}

From the extensive available data set and the good performance of the calorimeter in the test beam periods both at the PS and at the SPS we expect to perform a thorough testing of imaging hadron calorimetry with tungsten absorbers over the full energy range relevant for CLIC. The tests with scintillator readout will be complemented by the use of active layers  consisting of RPCs with digital readout with a granularity of $1\times1$ cm$^2$ in the tungsten structure in 2012.

\bibliographystyle{IEEEtran.bst}
\bibliography{CALICE}

\begin{thebibliography}{10}
\providecommand{\url}[1]{#1}
\csname url@rmstyle\endcsname
\providecommand{\newblock}{\relax}
\providecommand{\bibinfo}[2]{#2}
\providecommand\BIBentrySTDinterwordspacing{\spaceskip=0pt\relax}
\providecommand\BIBentryALTinterwordstretchfactor{4}
\providecommand\BIBentryALTinterwordspacing{\spaceskip=\fontdimen2\font plus
\BIBentryALTinterwordstretchfactor\fontdimen3\font minus
  \fontdimen4\font\relax}
\providecommand\BIBforeignlanguage[2]{{%
\expandafter\ifx\csname l@#1\endcsname\relax
\typeout{** WARNING: IEEEtran.bst: No hyphenation pattern has been}%
\typeout{** loaded for the language `#1'. Using the pattern for}%
\typeout{** the default language instead.}%
\else
\language=\csname l@#1\endcsname
\fi
#2}}

\bibitem{Brau:2007zza}
J.~Brau \emph{et~al.}, ``{International Linear Collider reference design
  report. 1: Executive summary. 2: Physics at the ILC. 3: Accelerator. 4:
  Detectors},'' \emph{ILC-REPORT-2007-001}.

\bibitem{Assmann:2000hg}
R.~Assmann, F.~Becker, R.~Bossart, H.~Burkhardt, H.-H. Braun, \emph{et~al.},
  ``{A 3-TeV {$e^+ e^-$} linear collider based on CLIC technology},''
  \emph{CERN-2000-008}, 2000.

\bibitem{Brient:2002gh}
J.-C. Brient and H.~Videau, ``{The Calorimetry at the future {$e^+e^-$} linear
  collider},'' \emph{arXiv:hep-ex/0202004}, 2001.

\bibitem{pfaMorgunov}
V.~L. Morgunov, ``{Calorimetry design with energy-flow concept (imaging
  detector for high-energy physics)},'' \emph{{CALOR 2002, Pasadena,
  California. Published in Pasadena 2002, 'Calorimetry in particle physics'}},
  2002.

\bibitem{Thomson:2009rp}
M.~Thomson, ``{Particle Flow Calorimetry and the PandoraPFA Algorithm},''
  \emph{Nucl.Instrum.Meth.}, vol. A611, pp. 25--40, 2009.

\bibitem{Adloff:2010hb}
C.~Adloff, Y.~Karyotakis, J.~Repond, A.~Brandt, H.~Brown, \emph{et~al.},
  ``{Construction and Commissioning of the CALICE Analog Hadron Calorimeter
  Prototype},'' \emph{JINST}, vol.~5, p. P05004, 2010.

\bibitem{Bondarenko:2000in}
G.~Bondarenko \emph{et~al.}, ``{Limited Geiger-mode microcell silicon
  photodiode: New results},'' \emph{Nucl. Instrum. Meth.}, vol. A442, pp.
  187--192, 2000.

\bibitem{Dyshkant:2006et}
A.~Dyshkant, ``{Tail Catcher Muon Tracker for the CALICE test beam},''
  \emph{AIP Conf. Proc.}, vol. 867, pp. 592--599, 2006.

\bibitem{Repond:2011jc}
J.~Repond, ``{Development of Particle Flow Calorimetry},''
  \emph{arXiv:1110.2121 [physics.ins-det]}, 2011.

\bibitem{Geant4Physics}
J.~Apostolakis \emph{et~al.}, ``Geant4 physics lists for {HEP},'' \emph{2008
  IEEE Nuclear Science Symposium Conference Record}, 2008.

\end{thebibliography}

\end{document}